\documentclass[a4paper,11pt]{article}

\usepackage[margin=1.14in]{geometry}
\usepackage{tabularx}
\usepackage{multirow}
\usepackage{booktabs}   
\usepackage{tabularx} 
\usepackage{arydshln}
\usepackage{amsmath}
\usepackage{amsthm}
\usepackage{amssymb}
\usepackage{graphicx}
\usepackage{pdfpages}
\makeatletter
\makeatletter \renewcommand{\@dotsep}{10000} \makeatother
\usepackage{wrapfig}
\usepackage[utf8]{inputenc}
\usepackage{lmodern}
\usepackage{hyperref}
\hypersetup{
	unicode,
	colorlinks,
	breaklinks,
	urlcolor=cyan, 
    linkcolor=black, 
	pdfauthor={Author One, Author Two, Author Three},
	pdfproducer={LaTeX},
	pdfcreator={pdflatex}
}

\definecolor{darkred}{rgb}{0.6, 0, 0}
\definecolor{darkblue}{rgb}{0, 0, 0.6}
\hypersetup{colorlinks=true, linkcolor=black, urlcolor=darkblue, citecolor=darkred}

\usepackage[T1]{fontenc} 
\usepackage{amsmath}

\newcommand{\be}{\begin{eqnarray}}
\newcommand{\ee}{\end{eqnarray}}
\def\be{\begin{equation}}
\def\ee{\end{equation}}
\def\bea{\begin{eqnarray}}
\def\eea{\end{eqnarray}}

\newcommand{\gsim}{\;\raisebox{-0.9ex}{$\textstyle\stackrel{\textstyle >}{\sim}$}\;}
\newcommand{\lsim}{\;\raisebox{-0.9ex}{$\textstyle\stackrel{\textstyle<}{\sim}$}\;}
\def\lsim{\raise0.3ex\hbox{$\;<$\kern-0.75em\raise-1.1ex\hbox{$\sim\;$}}}
\def\gsim{\raise0.3ex\hbox{$\;>$\kern-0.75em\raise-1.1ex\hbox{$\sim\;$}}}

\usepackage{graphics}

\usepackage{epsfig}
\usepackage{slashed}
\usepackage[utf8]{inputenc}
\usepackage{multirow}
\usepackage{pstricks}
\usepackage{dcolumn}

\usepackage[sort&compress,numbers,square]{natbib}

\theoremstyle{plain}

\theoremstyle{definition}

\usepackage{graphicx, color}

\title{Interference Effects in Resonant Standard Model di-Higgs Production  and Decay 
into $4b$ Final States: the Role of Machine Learning Analysis}
\vspace{10mm}
\author{\Large{A. Hammad$^{a}$\thanks{\href{mailto:ahammad@kek.jp}{ahammad@kek.jp}}\ ,  S. Moretti$^{b, c}$\thanks{\href{mailto:s.moretti@soton.ac.uk}{s.moretti@soton.ac.uk},\href{mailto:stefano.moretti@physics.uu.se}{stefano.moretti@physics.uu.se}}\ , A.P. Przybyl$^{c,d}$\thanks{\href{mailto:anna.przybyl@desy.de}{anna.przybyl@desy.de}}\   \ and H. Waltari$^c$}\thanks{\href{mailto:harri.waltari@physics.uu.se}{harri.waltari@physics.uu.se}}}

\date{
\small{
$^a$Theory Center, IPNS, KEK, 1-1 Oho, Tsukuba, Ibaraki 305-0801, Japan.\\
$^b$School of Physics and Astronomy, University of Southampton, Highfield,
Southampton, UK.\\
$^c$Department of Physics $\&$ Astronomy, Uppsala University, Box 516, SE-751 20 Uppsala, Sweden.\\
$^d$ Deutsches Elektronen-Synchrotron (DESY), Notkestrasse 85, 22607 Hamburg, Germany.
}}
\begin{document}
	\maketitle
	\vspace{4mm}
	\begin{abstract}
 \normalsize{The final state with four $b$-quarks has generally the largest event rate in Standard Model (SM)-like Higgs ($h_{\rm SM}$) pair production, but also the largest backgrounds. We study such a final state using the $gg\to h_{\rm SM}h_{\rm SM}$ production mechanism and Benchmarks Points (BPs) derived from the Next-to-Minimal Supersymmetric SM (NMSSM) in the boosted case, leading to two (fat) 'Higgs jets'. To suppress the backgrounds we use a combination of both kinematical cuts and jet substructure features exploiting Machine Learning (ML) analysis.
We simulate the signal BPs both with and without the interference of the resonant $s$-channel diagram with the non-resonant topologies emerging from both the SM and NMSSM. The ML architecture of choice here is based on a multi-modal Transformer, which performs significantly better than traditional ML algorithms, in two respects: firstly, it enables to achieve higher significances and, secondly,  it adapts better to the analysis dataset with interferences even if it was trained on one  without these. However, neglecting the effect of the latter in experimental searches could lead to grossly mistaken results.}
\end{abstract}
\newpage
\maketitle \flushbottom
\vspace{4mm}
\section{Introduction}

The discovery of Standard Model (SM) Higgs pair production is one of the main goals of the experimental collaborations at the Large Hadron Collider (LHC). At Leading Order (LO) Higgs pairs can be produced at one loop through the so called triangle and box diagrams with top quarks running in the loops. Higgs pair production is the most straightforward way to measure the trilinear Higgs self-coupling. It is also a probe of new physics including new colored particles in the loop \cite{Batell:2015koa,Huang:2017nnw,Moretti:2023dlx,DeCurtis:2023pus} or heavy Higgs bosons \cite{Plehn:1996wb,Dolan:2012ac,No:2013wsa,Dawson:2015haa} leading to modifications of non-resonant and resonant di-Higgs production, respectively. Recently also the importance of the interference effects in resonant di-Higgs production has been recognized in phenomenological studies \cite{Dawson:2016ugw,Carena:2018vpt,Feuerstake:2024uxs,DeCurtis:2023pus,Moretti:2025dfz}, while experimental searches typically assume a Breit-Wigner (BW) distribution\footnote{Neglecting the interference is justified if the new scalar is extremely narrow, but many realistic scenarios lead to significant widths, for which the interference is large.} \cite{ATLAS:2023vdy,CMS:2024pjq}.

A single Higgs boson has been observed in several decay channels ($b\bar{b}$, $WW^{*}$, $\tau\tau$, $ZZ^{*}$, $\gamma\gamma$ and evidence in $\mu\mu$), which means that the di-Higgs signal can be split into several possible final state topologies, each having their advantages and disadvantages. In this work we concentrate on the final state of four $b$-quarks, which has the highest event rate. The downside is that it also suffers from the highest background arising from QCD. Nevertheless, this channel is the most sensitive one for the SM signal hypothesis at high invariant masses \cite{ATLAS:2023vdy}, where the event rate is the lowest.

In the region of high invariant masses the Higgs bosons become boosted, so that individual $b$-jets cannot be resolved. Instead one will have to look at so-called fat jets, \textit{i.e.}, jets with a large jet radius containing the decay products of the Higgs boson \cite{ATLAS:2022hwc,CMS:2024pjq}. Such a, so to say, `Higgs jet' is thus characterized not only by a jet mass near the actual Higgs mass, 125 GeV or so, but also a substructure arising from the presence of two initial $b$-partons. For our chosen decay pattern, there would then be two such Higgs jets. 

The substructure patterns can be extracted from the tracks belonging to the fat jet and used to separate Beyond the SM (BSM)  signals from QCD backgrounds.  In this respect, it should be noted that, recently,
traditional methods have been surpassed by new techniques using advanced Machine Learning (ML) approaches  \cite{Chakraborty:2019imr,Chung:2020ysf,Guo:2020vvt,Khosa:2021cyk,Datta:2019ndh}. 
Furthermore, the study of kinematic  properties of such fat jets is also of relevance \cite{Cogollo:2020afo,Grossi:2020orx,Ngairangbam:2020ksz,Englert:2020ntw,Freitas:2020ttd,Stakia:2021pvp,Jorge:2021vpo,Ren:2021prq,Alvestad:2021sje,Jung:2021tym,Drees:2021oew,Cornell:2021gut,
Vidal:2021oed}. Furthermore, such a  kinematics ({\it i.e.}, encoding the global features of the final state jets), possibly together with the knowledge of the properties of possible new  intermediate particles is insensitive to the (local) dynamics occurring inside the jet themselves. Therefore, the global and local properties of the hadronic final state offer complementary insights into the underlying Higgs dynamics. 

A possible way to profitably combine information from both jet substructure and jet kinematics is to concatenate the two inputs through a multi-modal network, as done in Refs.~\cite{Lin:2018cin,Moreno:2019neq,Kim:2019wns,Chung:2022kjp,Huang:2022rne,Esmail:2023axd}. However, a simple concatenation leads to an imbalance of the extracted knowledge, in such a way that global kinematics generally dominates local substructures \cite{Ban:2023jfo}. Therefore, in
 a previous paper \cite{Hammad:2023sbd}, some of us presented a new method for incorporating information extracted from both global and local dynamics emerging from jets in a Transformer encoder with a cross-attention layer, extracting the most relevant information from each dataset individually using first self-attention layers. Such a method was proven to offer a sizable improvement in classification performance compared to the simple concatenation approach.

 It is the purpose of this paper to adopt such a ML approach  for the case of $gg\to H\to h_{\rm SM}h_{\rm SM}\to b\bar b b\bar b$ production and decay,  where $H$ is a heavy Higgs state with mass $m_H>2m_{h_{\rm SM}}$, but for a different modeling of the signal. While in  \cite{Hammad:2023sbd} a simplified  approach was adopted, whereby the box diagrams were neglected and the triangle ones where computed through factorization in Narrow Width Approximation (NWA) of the $H$ state, here, we retain both diagram topologies, crucially generating interference of the resonant diagram with both of the above. For comparison we also simulate the corresponding events with a pure BW description, \textit{i.e.}, without interferences. We do so as it has been demonstrated that such interferences can be significant,  {\it e.g.}, in the case of the Next-to-Minimal Supersymmetric SM (NMSSM) \cite{Moretti:2025dfz}, which we adopt here as illustrative example of an underlying Higgs scenario strongly affected  by such effects. We will show the phenomenological relevance of all such interferences,  normally neglected in experimental analysis, including highlighting  the response of our multi-modal Transformer with respect to the one induced by simpler ML methods. 

 The plan of the paper is as follows. In the next section we describe the Benchmark Points (BPs) used in the context of the NMSSM.  Then, in Sect.~\ref{sec:DL}, we discuss our ML infrastructure. Results of  our Monte Carlo (MC)  analysis and conclusions will follow in turn.

\section{The NMSSM Signal Scenarios}

We study here the BPs introduced in \cite{Moretti:2025dfz}, which were derived from the NMSSM spectrum enabling resonant (heavy) Higgs production and decay (see also Ref.~\cite{Moretti:2023dlx} for the non-resonant case). In \cite{Moretti:2025dfz} only a very basic event selection was performed and no backgrounds were considered. We aim now to estimate, what kind of prospects would a realistic analysis have to observe some of the BPs against the QCD background in the $4b$ final state.

For resonant di-Higgs production the most important part is the Higgs sector, which in the NMSSM consists of two Higgs doublets and a singlet. The scalar potential of the CP-even sector at tree-level has the form
\begin{equation}
\begin{split}\label{eq:scalarpotential}
V(H_{u}^{0},H_{d}^{0},S) &= \frac{1}{2}m_{H_{u}}^{2}(H_{u}^{0})^{2}+\frac{1}{2}m_{H_{d}}^{2}(H_{d}^{0})^{2}+\frac{1}{2}m_{S}^{2}S^{2}+\frac{A_{\lambda}}{2\sqrt{2}}SH_{u}^{0}H_{d}^{0}+\frac{A_{\kappa}}{6\sqrt{2}}S^{3}\\
&+\frac{1}{32}(g^{2}+g^{\prime 2})((H_{d}^{0})^{2}-(H_{u}^{0})^{2})^{2}
+\frac{1}{8}g^{2}(H_{d}^{0}H_{u}^{0})^{2}
\\& +\frac{\lambda^{2}}{4}(|H_{u}^{0}H_{d}^{0}|^{2}+|H_{u}^{0}S|^{2}+|H_{d}^{0}S|^{2})
+\frac{\kappa\lambda}{4}H_{u}^{0}H_{d}^{0}S^{2}+\frac{\kappa^{2}}{4}S^{4}.
\end{split}
\end{equation}

Here $g$ and $g^{\prime}$ are the electroweak gauge couplings, $\lambda$ and $\kappa$ are the superpotential couplings between the three Higgs superfields and the cubic singlet term, respectively and $A_{\lambda}$ and $A_{\kappa}$ are couplings for the corresponding soft supersymmetry breaking scalar interactions.

Furthermore, we add a quartic term arising from the top-stop loops, which we parametrize in the form
\begin{equation}
    V_{\mathrm{loop}}=\delta |H_{u}^{0}|^{4}.
\end{equation}

We assume here that the 125 GeV Higgs boson is the lighter doublet scalar. In general the heavy doublet scalar has large couplings to third generation fermions, while the fermionic couplings of the singlet are induced by mixing with the doublets. Since the mixing is small unless the states are close in mass, the singlet is usually a narrow resonance with $\Gamma/m\sim 10^{-3}$ while the heavy doublet has a larger width $\Gamma/m \sim 10^{-2}$. Trilinear couplings with vector bosons are nearly always small for both BSM Higgs states.

The resonant di-Higgs production proceeds through top and bottom quark loops. For the doublets the coupling to quarks is of the order of the largest Yukawa couplings, while for the singlets it is controlled by the mixing with doublets. The event rates also depend on the trilinear Higgs couplings, which read in the alignment limit
\begin{eqnarray}
    \lambda_{Hhh} & = &\left(\frac{3(g^{2}+2\lambda^{2})}{16}\sin 4\beta + 12\delta \sin^{3}\beta\cos\beta\right) v, \label{eq:Hhhcoupling}\\
    \lambda_{Shh} & = & \frac{\lambda^{2}}{2}v_{S}+\left(\frac{A_{\lambda}}{4\sqrt{2}}+\frac{\kappa\lambda}{4}v_{S}\right)\sin 2\beta.
\end{eqnarray}
Here $v=246$~GeV is the vacuum expectation value (VEV) of the SM Higgs and $v_{S}$ is the VEV of the singlet. Equation (\ref{eq:Hhhcoupling}) also shows why we are not considering the MSSM: in the MSSM a high value of $\tan\beta$ is required to achieve a $125$~GeV Higgs and at high $\tan\beta$ this expression goes to zero suppressing any resonant di-Higgs production. The parameter region with the largest resonant di-Higgs signal is that of moderate $\tan\beta$ and large $\lambda$.

ATLAS performs di-Higgs searches in the $4b$ channel in both the resolved \cite{ATLAS:2023qzf} and boosted \cite{ATLAS:2024lsk} case. The limits for the boosted case were derived for $m_{H}=1\ldots 5$~TeV, but the onset of the boosted signal starts already at slightly lower invariant masses. Therefore we analyse BPs starting from $m_{H}=800$~GeV, which has a higher overall event rate than the BP with $m_{H}\geq 1$~TeV, but for which the boosted selection efficiency is not yet maximal.

We set up two BPs with opposite interference patterns having heavy scalars around $800$~GeV. The most important parameters for the BPs are given in table \ref{tab:parameters}. These are BP4 and BP5 of \cite{Moretti:2025dfz}, where the effects of squarks and the singlet scalar have been left out.  BP1 has destructive interference when $m_{hh}<m_{H}$ and constructive interference when $m_{hh}>m_{H}$. This leads to an overall reduction of the cross section as the differential cross section of the continuum is larger in the region of destructive interference. BP2 has the opposite pattern, constuctive interference when $m_{hh}<m_{H}$ and destructive when $m_{hh}>m_{H}$. This leads to a slight enhancement of the cross section compared to the case without the interference.

\begin{table}[htb]
    \centering
\begin{tabular}{c|r r}
Parameter & BP1 & BP2\\
\hline\hline
$\tan\beta$ & $2.31$ & $7.00$\\
$\lambda$ & $0.65$ & $0.21$\\
$\kappa$  & $0.68$ & $0.16$\\
$A_{\lambda}$ (GeV) & $220$ & $-550$\\
$v_{s}$ (GeV) & $1280$ & $943$\\
$m_{H}$ (GeV) & $800$ & $845$\\
$\Gamma_{H}$ (GeV) & $13.3$ & $2.4$\\
$\sigma (pp\rightarrow hh)$ (fb) & $14.9$ & $18.3$\\
$\sigma (pp\rightarrow hh)_{BW}$ (fb) & $17.8$ & $17.4$
\end{tabular}
    \caption{The most important parameters for the BPs and the LO cross sections for di-Higgs production with and without interferences of the resonant production with the SM-like di-Higgs continuum.}
    \label{tab:parameters}
\end{table}

\begin{figure*}[!h]
    \includegraphics[width=0.95\linewidth]{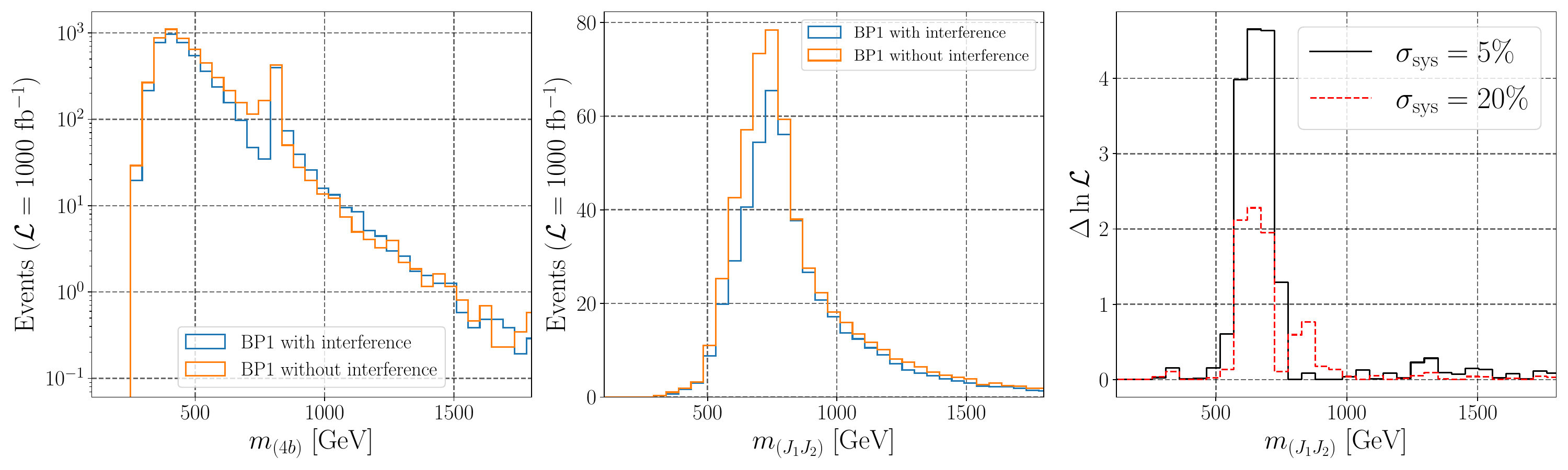} \\
    \includegraphics[width=0.95\linewidth]{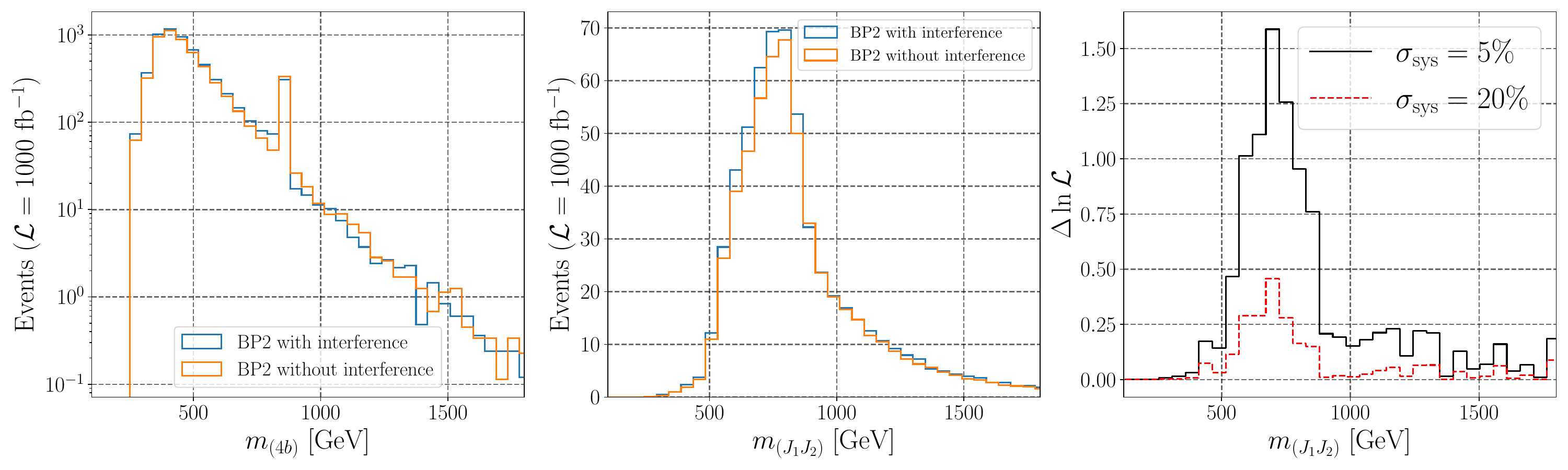}
    \caption{Left: Invariant mass distribution of the heavy mediator reconstructed at parton level, normalized to the expected number of events at an integrated luminosity of $1000~\mathrm{fb}^{-1}$.
Middle: Invariant mass distribution of the heavy mediator reconstructed from the fat jet final state, normalized to the expected number of events at an integrated luminosity of $1000~\mathrm{fb}^{-1}$.  
Right: Poisson likelihood per bin quantifying the difference between the two distributions shown on the left, including systematic uncertainties of $5\%$ (black) and $20\%$ (red).
}
    \label{fig:invmass}
\end{figure*}

The left panels of figure \ref{fig:invmass} show the impact of interference effects for a heavy doublet scalar. At parton level the destructive interference for BP1 can result locally up to a $70\%$ reduction in the differential cross section. Detector effects will smear this out somewhat, but an overall reduction in the event rate can be substantial. For BP2 the effect is not as large, since the decay width of the heavy scalar is smaller. Even in this case the interference is not negligible.

To quantify the difference between two histograms bin by bin, we utilize the Poisson likelihood ratio. Let $O_i$ be the observed number of events in bin $i$ ({\it e.g.}, from signal with interference) and $E_i$ the expected number of events ({\it e.g.}, from signal without interference). Assuming Poisson statistics, the per-bin likelihood ratio is defined as 
\begin{equation}
\Delta\ln\mathcal{L}_i = 2 \left[ E_i - O_i + O_i \ln \frac{O_i}{E_i} \right],
\end{equation}
where the logarithm is taken to be zero if $O_i=0$. The total test statistic is obtained by summing over all bins, $\Lambda = \sum_i \Delta\ln\mathcal{L}_i$, which asymptotically follows a $\chi^2$ distribution with degrees of freedom equal to the number of bins. Systematic uncertainties can be included by combining them in quadrature with the statistical uncertainty: if $\sigma_{{\rm sys},i}$ denotes the systematic uncertainty in bin $i$, the total uncertainty becomes $\sigma_i = \sqrt{E_i + \sigma_{{\rm sys},i}^2}$. An approximate way to include the systematic effect in the Poisson likelihood is to replace $E_i$ by $E_i + \sigma_{{\rm sys},i}$ in the likelihood ratio, giving $\Delta\ln\mathcal{L}_i= 2 \left[ (E_i + \sigma_{{\rm sys},i}) - O_i + O_i \ln \frac{O_i}{E_i + \sigma_{{\rm sys},i}} \right]$. A more rigorous treatment introduces the systematic as a nuisance parameter in a profile likelihood. This approach provides a per-bin statistical measure of the difference between the two distributions while accounting for both statistical and systematic uncertainties. The right panels of figure \ref{fig:invmass} shows the  Poisson likelihood for uncertainty of $5\%$ (black) and $20\%$ (red) for the two signal scenarios. 


\section{ML Analysis}
\label{sec:DL}
In addition to global kinematic observables, jet substructure information provides a powerful means of distinguishing signal from background events. 
This stems from the fact that jets initiated by different parent particles exhibit distinct internal patterns.  For example, heavy boosted particles such as the $W^{\pm}$, $Z$ and Higgs bosons typically decay into multiple partons, producing jets with a multi-prong structure. 
In contrast, quark- and gluon-initiated QCD jets usually display a simpler, single pronge structure.  These differences imply that the features of the parent particle can be inferred from the internal organization of jet constituents.  By exploiting this substructure, one can recover localized event information that is otherwise inaccessible through kinematic observables alone, making jet substructure a key discriminant between processes of different physical origins.  The concept of utilizing ML methods to identify the particle initiating a jet, and thereby distinguish jets from boosted heavy objects and those from QCD processes, was  first introduced in~\cite{Cogan:2014oua,Almeida:2015jua,deOliveira:2015xxd,Baldi:2016fql,Barnard:2016qma,Komiske:2016rsd,Kasieczka:2017nvn,Macaluso:2018tck,Choi:2018dag,Guest:2016iqz,Pearkes:2017hku,Egan:2017ojy,Fraser:2018ieu,Butter:2017cot,Kasieczka:2018lwf,Esmail:2025kii,Hammad:2024cae,Hammad:2024hhm}. 

Building on these methods, we utilize a multi-modal Transformer encoder with multi-head self-attention (hereafter, `Transformer' for short)  to analyze jet contents that has already been introduced in \cite{Hammad:2023sbd}. 
The model architecture incorporates three distinct processing streams designed to capture information at different energy levels. 
The first and second streams independently analyze the constituents of the leading and sub-leading Higgs jets, each employing a Transformer encoder with self-attention layers.  After extracting the salient jet features, their outputs are combined through an addition layer. The third stream processes the high-level kinematic variables, using another Transformer encoder with self-attention heads. 
Finally, the jet-based and kinematic features are merged via a cross-attention layer, which allows the network to learn correlations between the jet substructure and global event kinematics.

\subsection{Data Preprocessing}
We begin by preprocessing the data sets corresponding to the leading and second-leading jets, each containing up to $50$ constituents. 
The constituents are ordered in descending transverse momentum ($p_T$), and for events with fewer than $50$ particles, the remaining entries are padded with zeros to ensure a uniform input size. 
An attention mask is applied so that the network performance remains unaffected by the zero-padded entries~\cite{vaswani2017attention}. 

For each jet constituent, we store seven input features \cite{Qu:2019gqs}: 
\begin{equation}
    \begin{aligned}
    &\text{- } \Delta\eta = \eta - \eta_{\text{jet}} && \text{: pseudorapidity difference} \\
    &\text{- } \Delta\phi = \phi - \phi_{\text{jet}} && \text{: azimuthal angle difference} \\
    &\text{- } \Delta R = \sqrt{(\Delta\eta)^2 + (\Delta\phi)^2} && \text{: angular distance from jet axis} \\
    &\text{- } \log(p_T) && \text{: transverse momentum (GeV) \; \; \; \; \; \; \; \; \; \; \;\;\; \; \; \; \; \; \;} \\
    &\text{- } \log(E) && \text{: energy (GeV)} \\
    &\text{- } \log\left(\frac{p_T}{p_{T_{\text{jet}}}}\right) && \text{: normalized $p_T$ (GeV)} \\
    &\text{- } \log\left(\frac{E}{E_{\text{jet}}}\right) && \text{: normalized energy (GeV)}
    \end{aligned}
    \label{eq:feature_matrix}
\end{equation}

Proper preprocessing of the jet contents is crucial for enhancing the discriminative capability of the network, especially for highlighting the multi-prong substructure characteristic of signal events. 
We adopt the preprocessing strategy originally developed for jet image analysis, which allows the model to learn effectively from relatively small datasets and significantly accelerates convergence during training.In principle, one may train the network directly on raw inputs, but this requires a substantially larger dataset and longer training times. Accordingly, we preprocess the input datasets for and efficient network training. 

The following geometric transformations are applied to each jet prior to network input.

\begin{itemize}
    \item \textbf{Translation:}  
    Jet constituents are shifted in the $\eta$--$\phi$ plane such that the jet axis is centered at the origin.

    \item \textbf{Rotation:}  
    To reduce stochastic variations due to random decay orientations, each jet is rotated to align its principal axis vertically.  
    This is achieved by computing the covariance matrix of the constituent positions in the $(\eta,\phi)$ plane and determining its leading eigenvector.  
    The rotation angle is defined as $\theta = \mathrm{arctan}(x_1, x_2)$, where $(x_1, x_2)$ are the components of the principal eigenvector.  
    The jet constituents are then rotated to new coordinates $(\eta', \phi')$, ensuring consistent alignment across events. 

    \item \textbf{Flipping:}  
    Finally, jets are reflected about the vertical axis so that the region with the highest momentum is always positioned on the right-hand side of the $\eta'$ axis.  
    This ensures that the hardest radiation appears consistently in the same region, improving classification performance.
\end{itemize}

After these preprocessing steps, the jet constituent data for the leading and second-leading jets acquire the dimensions $(n, 50, 7)$, where $n$ is the number of events, $50$ the number of constituents and $7$ the number of features per constituent.

\begin{figure*}[!h]
    \centering
    \includegraphics[width=0.9\linewidth]{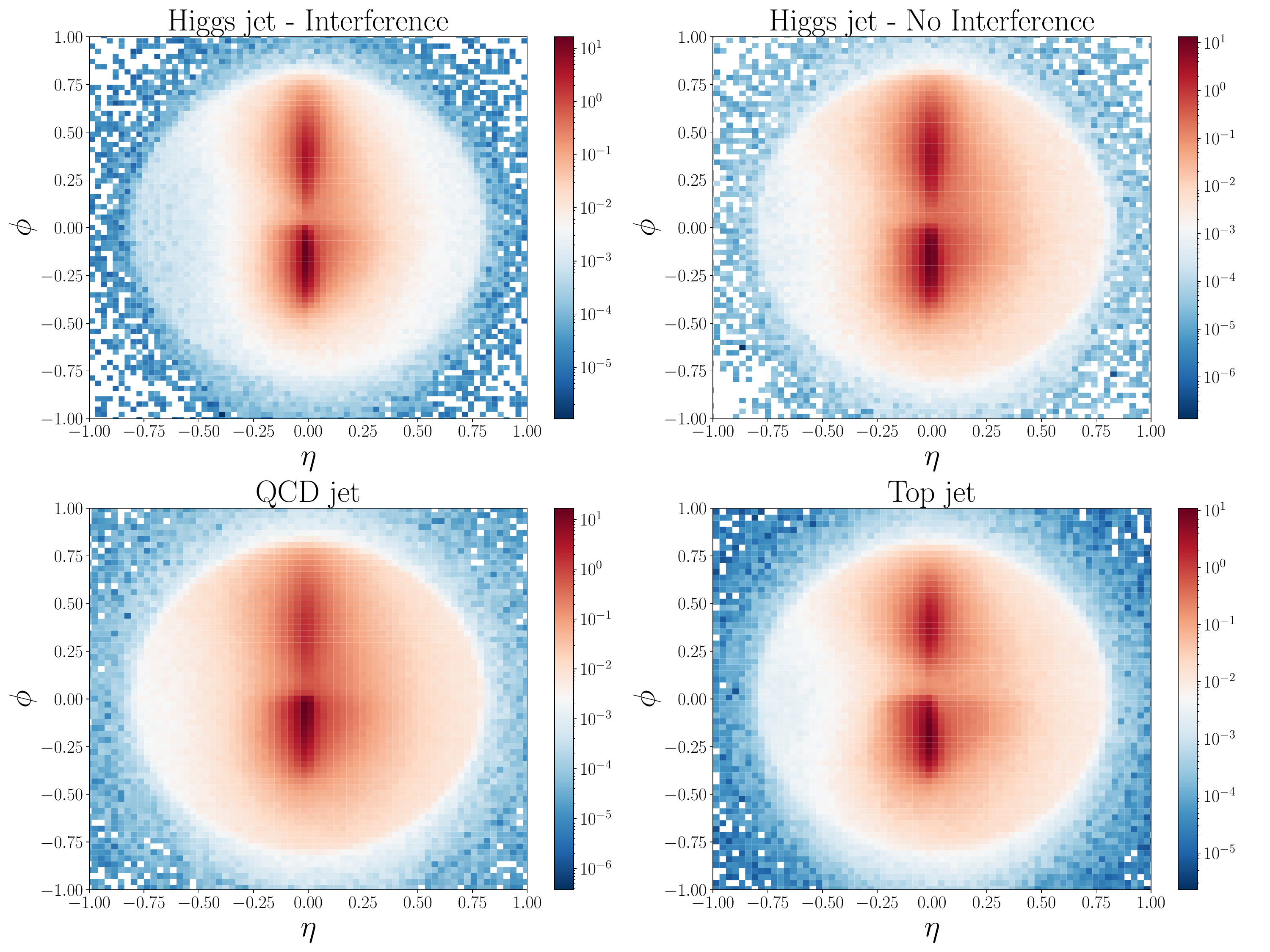}
    \caption{Accumulated average transverse momentum of hadrons for the leading jet, after preprocessing, computed over 50,000 events. }
    \label{fig:jet_images}
\end{figure*}

Figure~\ref{fig:jet_images} shows the cumulative average of $5\times10^4$ transverse momentum distributions for the leading Higgs jets for signal events (upper row) and background events (bottom row), after applying all the selection cuts. 
The preprocessing transformations clearly reveal the multi-prong structure characteristic of signal events, where subjets appear localized in distinct regions of the $(\eta, \phi)$ plane.  
In contrast, QCD multi-jet backgrounds exhibit a broad, featureless energy distribution without a clear prong structure, whereas $t\bar{t}$ events display a three-prong topology associated with hadronic top decays.  
Although $t\bar{t}$ events represent only about $10\%$ of the background sample, their distinct structure plays a useful role in evaluating the signal discriminability.

In addition to constituent-level information, we include high-level kinematic variables describing the reconstructed objects: the leading jet, second-leading jet, and heavy Higgs candidate.  
The corresponding kinematic dataset has dimensions $(n,3,5)$, where the five features per reconstructed object are:
\[
\{m, p_T, \eta, \phi, E \}\,.
\]

\begin{figure*}[!h]
    \centering 
    \includegraphics[width=0.9\linewidth]{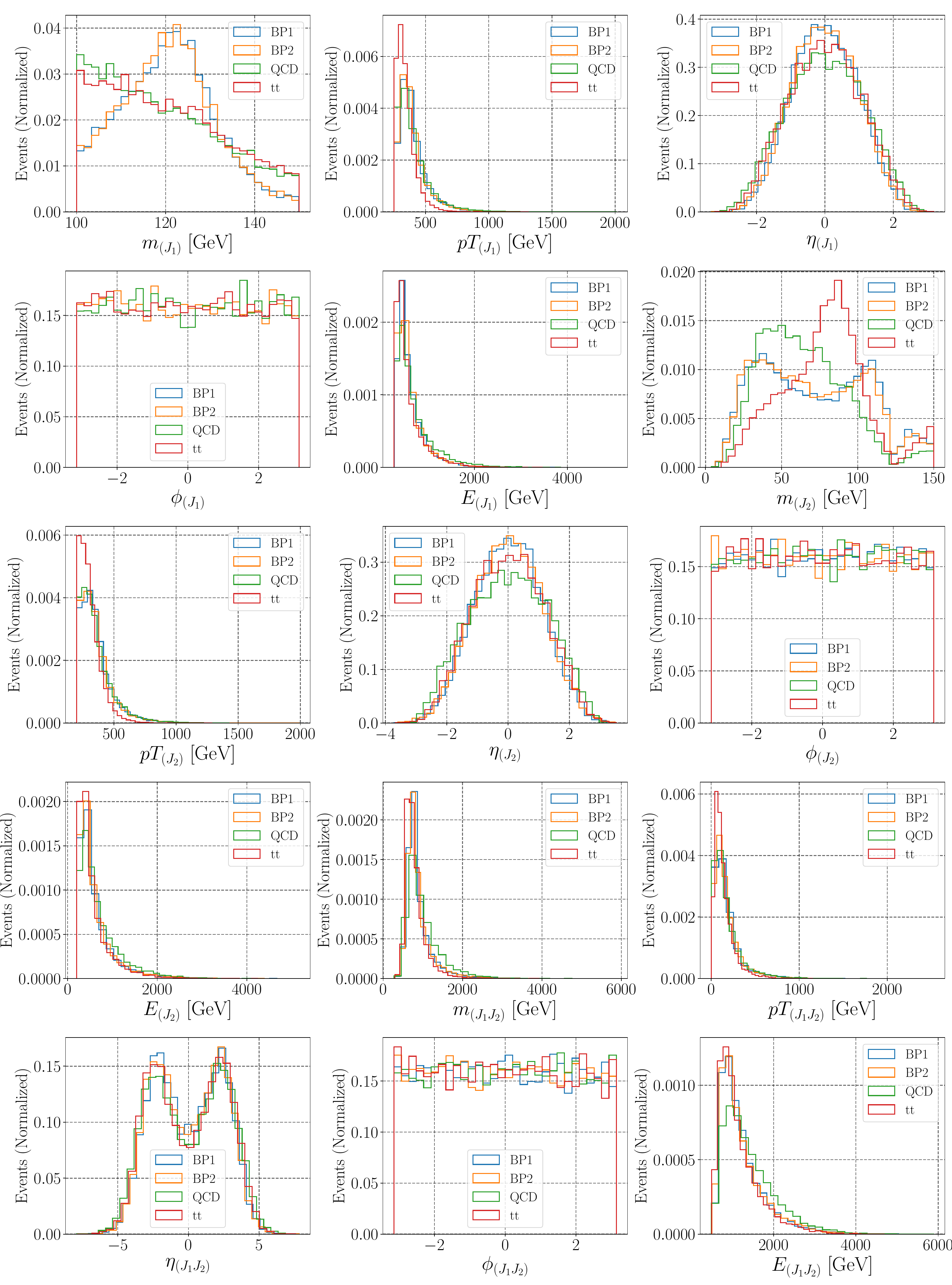}
    \caption{Kinematic distributions used for training. }
    \label{fig:kinematics}
\end{figure*}

Figure~\ref{fig:kinematics} displays the normalized kinematic distributions for the signal benchmarks and for the main background processes.  
We also include the rotation angles $\theta_i$ of the leading and second-leading jets (but not the Higgs candidate) to capture correlations between jet orientation and event topology.

For the Transformer network, the processed datasets are fed into three Transformer encoders.  
The first and second encoders receive inputs of shape $(n,50,7)$ corresponding to the leading and second-leading jets, respectively, while the third encoder processes the kinematic inputs of shape $(n,3,5)$.  
Signal and background events are stacked into separate datasets, labeled $Y=1$ for signal and $Y=0$ for background.  

For comparison we take a Multi-Layer Perceptron (MLP) architecture, which is a type of artificial Neural Network  
(NN) with one input layer, (at least) one hidden layer and one output layer, connected by weighted nodes, a setup which is typical of experimental approaches at the LHC. For its inputs the  kinematic distributions are reshaped into $(n,15)$, where $n$ is the number of training events with $15$ kinematic variables. Therefore the MLP input layers is adopted with $15$ neurons, corresponding to the size of the input distributions. Importantly, during the structure of the MLP layers we use only the kinematic information while it is not feasible how to include the jet-substructure information into a single network.

During training, the model minimizes a categorical cross-entropy loss function, optimizing the difference between predictions and true labels.  
We employ balanced datasets of equal size for signal and background, with $10^6$ events used for training and $10^5$ events reserved for testing.\footnote{A key characteristic of attention-based Transformer models is that their classification performance generally improves with the size of the training dataset.}

\subsection{Signal and Background Events Generation }

The primary background contamination in this analysis arises from QCD multijet processes, specifically $pp \to jjjj$, which contribute approximately $90\%$ of the total background. The di-top process $pp \to t\bar{t}$ contributes at the $10\%$ level, while other background processes, including SM $h$, $hh$, and EW di-boson production, have been found to make negligible contributions to the selected event yields and are therefore not included in our analysis.


For event generation, we use \textsc{MadGraph5}~\cite{Alwall:2014hca} to compute multi-parton amplitudes and to simulate both signal and background processes. The QCD background $pp \rightarrow b\bar{b}b\bar{b}$ and the $pp \rightarrow t\bar{t}$ background are computed at Leading Order (LO). The di-Higgs production via gluon-gluon fusion is calculated at one-loop, which is LO for this process, using the UFO model file from \cite{ResonantUFOmodel}. In the signal generation only the SM contribution, the heavy Higgs contribution and their interference were simulated. Given the experimental constraints on squarks \cite{CMS:2021eha,ATLAS:2024lda}, any squark entering the loop would be off-shell at $m_{hh}=m_{H}$ so the contribution around the resonance would be minimal and hence neglecting squarks in the production of the heavy Higgs is justified. The couplings were calculated with a private modification of \textsc{SPheno}~\cite{Porod:2003um,Porod:2011nf}. 
\textsc{PYTHIA}~\cite{Sjostrand:2006za} is employed for parton showering, hadronization, heavy-flavor decays, and the inclusion of the soft underlying event. The $t\bar{t}$ background is further simulated at LO with up to two additional jets using the MLM matching scheme~\cite{Alwall:2007fs,Mangano:2006rw} with a matching scale of $20~\mathrm{GeV}$.

\subsection{Cut-based Analysis}

The analysis began by applying a set of preselection cuts to the generated events. The resonance is heavy, so fat jet analysis is considered, with $hh \rightarrow b \bar{b} b \bar{b}$ being reconstructed as two fat jets. A Higgs tagging was performed, where the fat jet closest (next closest) to the Higgs mass was labelled as `fat jet 1' (`fat jet 2'). Events were required to have at least two fat jets, and additional cuts were placed on total hadronic transverse momentum (THT), as well as jet masses and transverse momenta ($p_T$). A summary of the cuts applied can be found in table \ref{tab:preselection_cuts}.

\begin{table*}[!ht]
    \centering
    \renewcommand{\arraystretch}{1.3}

    \begin{tabular}{l p{2.cm} p{2.8cm} p{2.7cm} p{2.8cm} p{2.2cm}}
        \toprule
        & \textbf{THT}
        & \textbf{Fat Jet 1 Mass}
        & \textbf{Fat Jet 1 $p_T$}
        & \textbf{Fat Jet 2 Mass}
        & \textbf{Fat Jet 2 $p_T$} \\
        \midrule
        $N_b \ge 2$ 
        & THT $> 300$
        & $100 < m < 150$
        & $p_T > 250$
        & $m < 150$
        & $p_T > 200$ \\
        \bottomrule
    \end{tabular}

    \caption{Preselection cuts applied to the resonant di-Higgs benchmark points. All units are in GeV.}
    \label{tab:preselection_cuts}
\end{table*}

A fully hadronic event arising from a heavy Higgs boson should contain a lot of hadronic activity and therefore we require $THT>300$~GeV. As a result of the Higgs tagging, the first fat jet mass has a well-defined peak very close to 125 GeV. Therefore, a cut was placed such that the mass had to be between 100 and 150 GeV. The second fat jet mass had a less defined peak and the peak itself was at a lower mass. This difference is reflected in the cut, with only an upper bound placed on this jet. Many signal events were killed when attempting to introduce a lower bound.
\begin{figure}[!ht]
    \includegraphics[width=0.95\linewidth]{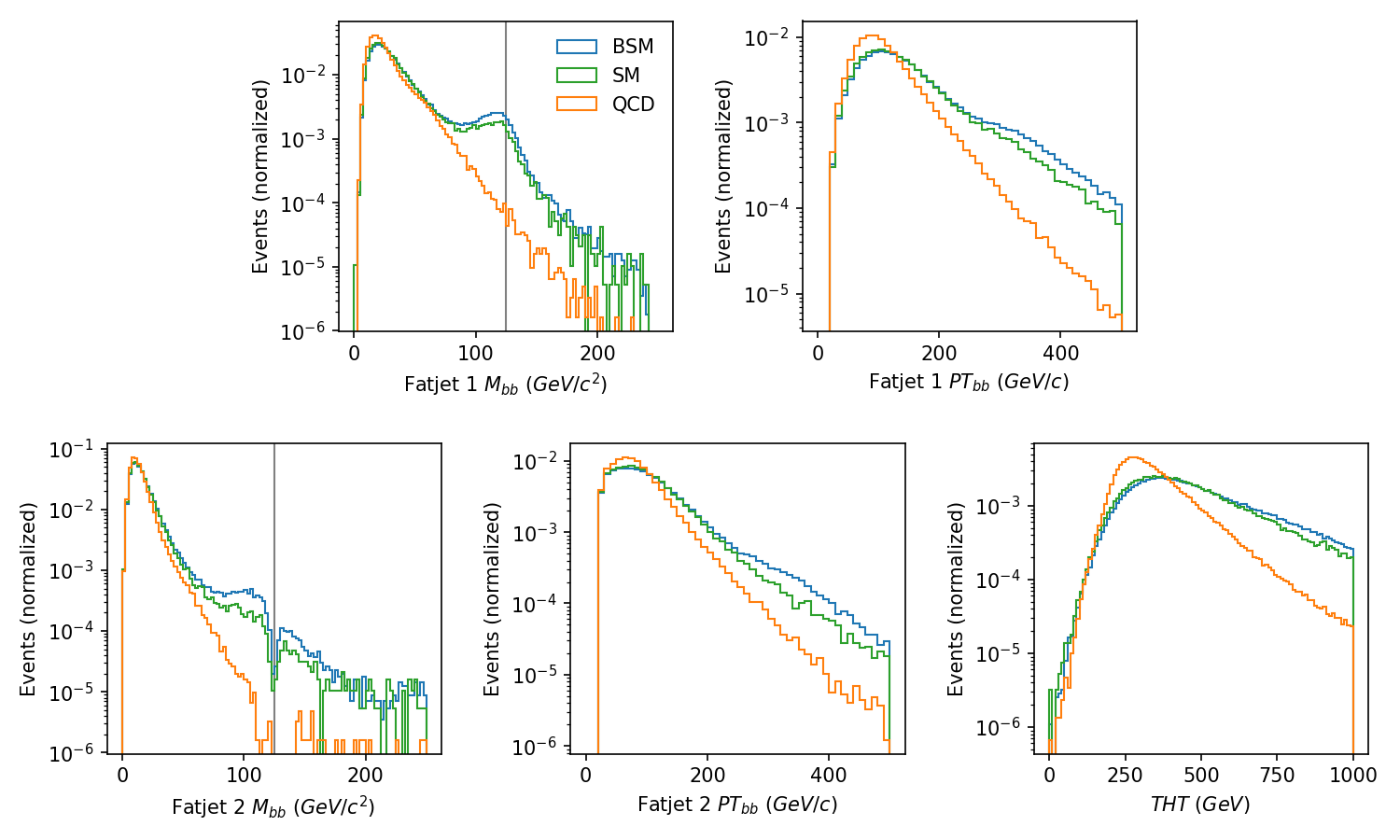}
    \caption{Kinematic variables used in the preselection cuts.}
    \label{fig:preselection_cuts}
\end{figure}
The BPs have a sharp resonant peak at $M_{4b}\simeq 800$~GeV at the parton level. At the detector level, the peak spreads out and shifts to the left, but a BSM di-Higgs excess can still be seen. The previous cuts reduced mainly the QCD background, but the final cuts on jet $p_T$ focus on emphasizing the BSM di-Higgs excess. The kinematics for the BSM di-Higgs and SM di-Higgs processes are similar, so the $p_T$ curves follow each other closely. There is a point where the $p_T$ begins to deviate for the BSM di-Higgs events compared to the SM di-Higgs events, so this was the region chosen to define the $p_T$ cuts. The BSM di-Higgs excess (as compared to the SM di-Higgs events) becomes more defined in the signal region as a result of the application of preselection cuts.

\begin{figure}
    \includegraphics[width=0.95\linewidth]{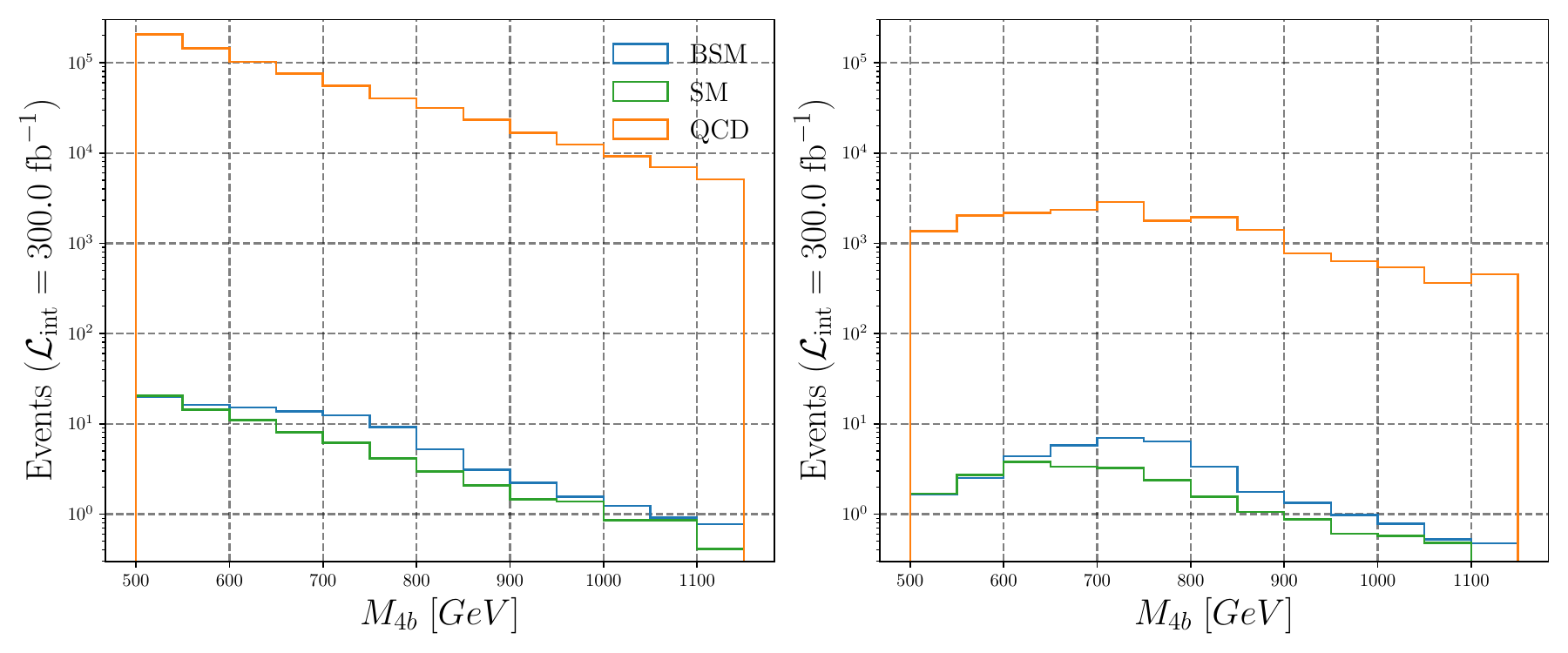}
    \caption{The $4b$ invariant mass before (left) and after (right) the preselection cuts.}
    \label{fig:4bmass_preselection}
\end{figure}

\begin{table}[!ht]
    \centering
    \renewcommand{\arraystretch}{1.25}
    \setlength{\tabcolsep}{12pt}

    \begin{tabular}{lcc}
        \toprule
        \textbf{Process} & \textbf{Before Cut} & \textbf{After Cut} \\
        \midrule
        BSM di-Higgs & 66 & 33 \\
        SM di-Higgs  & 39 & 18 \\
        Signal       & 27 & 15 \\
        QCD 4b       & 383{,}012 & 15{,}401 \\
        \midrule
        $\displaystyle \frac{S}{\sqrt{B}}$ & 0.0426 & 0.118 \\
        \bottomrule
    \end{tabular}

    \caption{Signal and background events in the signal region before and after the preselection cuts. A generator-level cut on the total transverse momentum was applied for the QCD background. The integrated luminosity is $\mathcal{L}_\text{int}=300~\mathrm{fb}^{-1}$.}
    \label{tab:preselection_events}
\end{table}

The number of events before and after the preselection cuts are applied can be seen in table \ref{tab:preselection_events}. The signal S is defined as the difference in the number of BSM di-Higgs events and SM di-Higgs events. The background B is the sum of QCD $4b$ background events and SM di-Higgs events. The signal significance can be estimated by $\frac{\text{S}}{\sqrt{\text{B}}}$, and can be found in table \ref{tab:preselection_events}. The selection efficiency $\epsilon$ is defined by $n_f/n_i$, where $n_i$ is the initial number of events, and $n_f$ is the final number of events after the preselection cuts are applied. The selection efficiencies that the mentioned cuts yield are $\epsilon_{\text{S}} = 55.6\%$ and $\epsilon_{\text{B}} = 4.0\%$, but it is to be noted that a generator level cut on THT was applied to the background. Hence the actual background rejection probability is better. The values used in the cuts were first chosen from the kinematic variable plots (see figure \ref{fig:preselection_cuts}), and later optimized to achieve the highest signal significance possible. We may notice that without ML methods the significance of the signal in the $4b$ channel remains low.

\section{Results}
Once the datasets are prepared, the NNs are trained to capture the non-linear correlations between the input features and their corresponding class labels. As described earlier, signal events are assigned the label $Y = 1$ while background events are assigned $Y = 0$. To eliminate any dependence of the training upon the ordering of the samples, signal and background events are merged into a single dataset and randomly shuffled together with their labels. Training is carried out in epochs, where one epoch is defined as a complete pass over the entire training dataset. During each epoch, the NN updates its trainable parameters through back-propagation, iteratively adjusting the weights to reduce the discrepancy between the predicted and true labels. The optimization procedure aims to minimize a chosen loss function by approaching its global minimum. For the architectures considered in this work, the NNs are trained for 30 epochs using a batch size of 256. The final output of each model is a probability vector $\hat{Y}$ of dimension $1 \times 2$, $ \hat{Y} = (\mathcal{P}_{\mathrm{sig}},\, \mathcal{P}_{\mathrm{bkg}})$,
where both components lie within the interval $[0,1]$. An event is classified as signal if $\mathcal{P}_{\mathrm{sig}} > 0.5$ (equivalently, $\mathcal{P}_{\mathrm{bkg}} < 0.5$), and as background if $\mathcal{P}_{\mathrm{sig}} < 0.5$ (or $\mathcal{P}_{\mathrm{bkg}} > 0.5$).

\begin{figure*}[!h]
    \includegraphics[width=\linewidth]{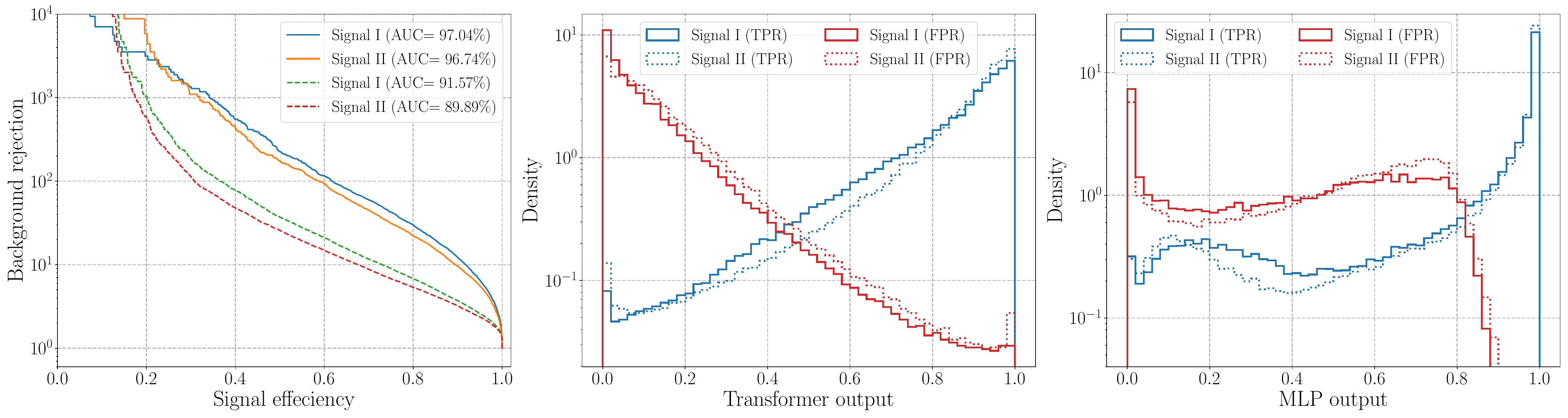}\\
    \includegraphics[width=\linewidth]{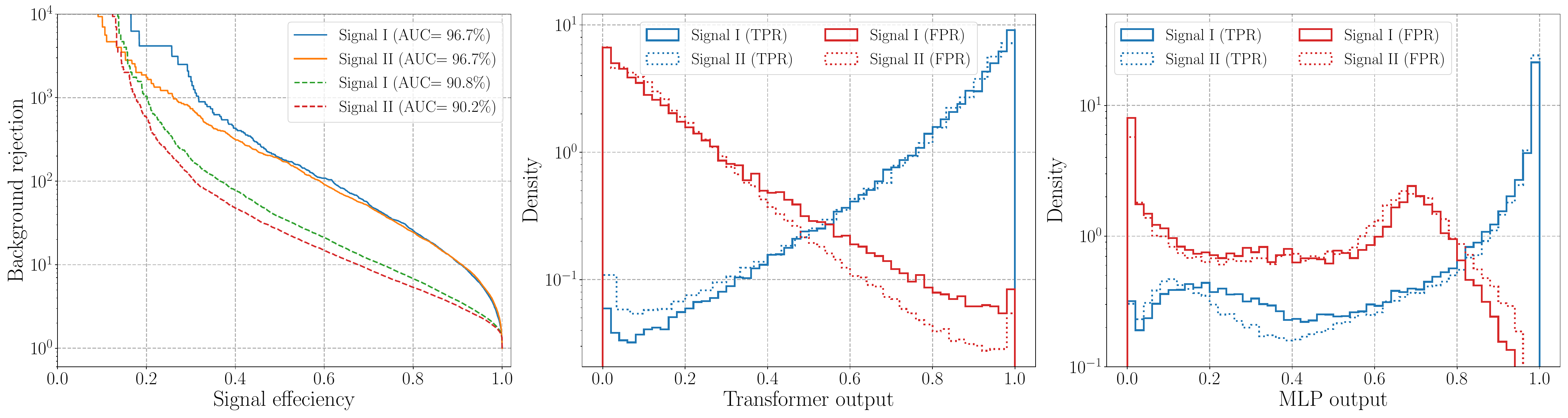}
    \caption{Signal I: NN trained on signal+interference and tested on the same dataset.     
    Signal II: NN trained on resonance only and tested on the  signal+interference dataset. }
    \label{fig:AUC}
\end{figure*}

In this analysis, we consider two training strategies. The first scenario, Signal I, involves training and testing the NN on signal events that include interference effects. This corresponds to the physically accurate situation of the NMSSM being realized in Nature. The second scenario, Signal II,  trains the network on signal events without interference but evaluates it on events with interference. This setup reflects the procedure commonly used in experimental analyses, in which the NN is trained on an idealized resonance signal sample but applied to real data that inherently contains interference effects. We then test the HL-LHC sensitivity for the two BPs in table \ref{tab:parameters}, with the two mentioned scenarios, using both the MLP and Transformer. 

The discriminating power of each network reflects how effectively it can separate signal from background events by exploiting their distinct underlying characteristics, which are intertwined across various kinematic distributions and jet structures. Figure \ref{fig:AUC} illustrates the Transformer (solid lines) and MLP (dashed lines) outputs, together with the corresponding Receiver Operating Characteristic (ROC) curves, for BP1 in the upper row and BP2 in the lower row. In both BPs, we compare the two training strategies using the correctly and incorrectly constructed training datasets. For BP1, where the negative interference significantly distorts the invariant mass distribution of the $4b$, the Transformer demonstrates greater robustness than the MLP. In contrast, for BP2, where the positive interference does not alter the shape of the invariant mass distribution, both the Transformer and the MLP show comparable sensitivity across the two training scenarios, although the Transformer still achieves notably better overall classification performance.
\begin{figure*}[!h]
    \centering
    \includegraphics[width=0.49\linewidth]{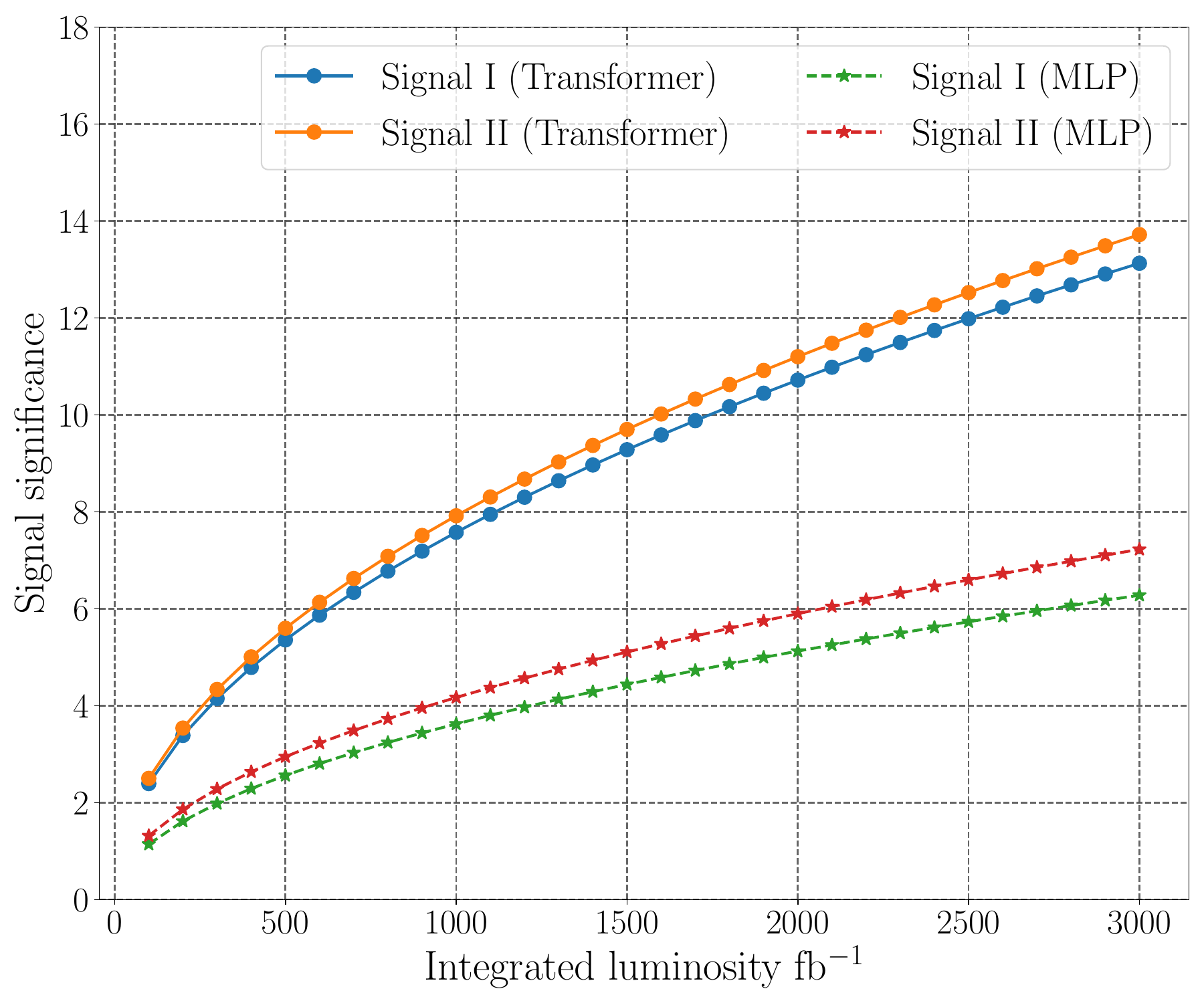}~~
    \includegraphics[width=0.49\linewidth]{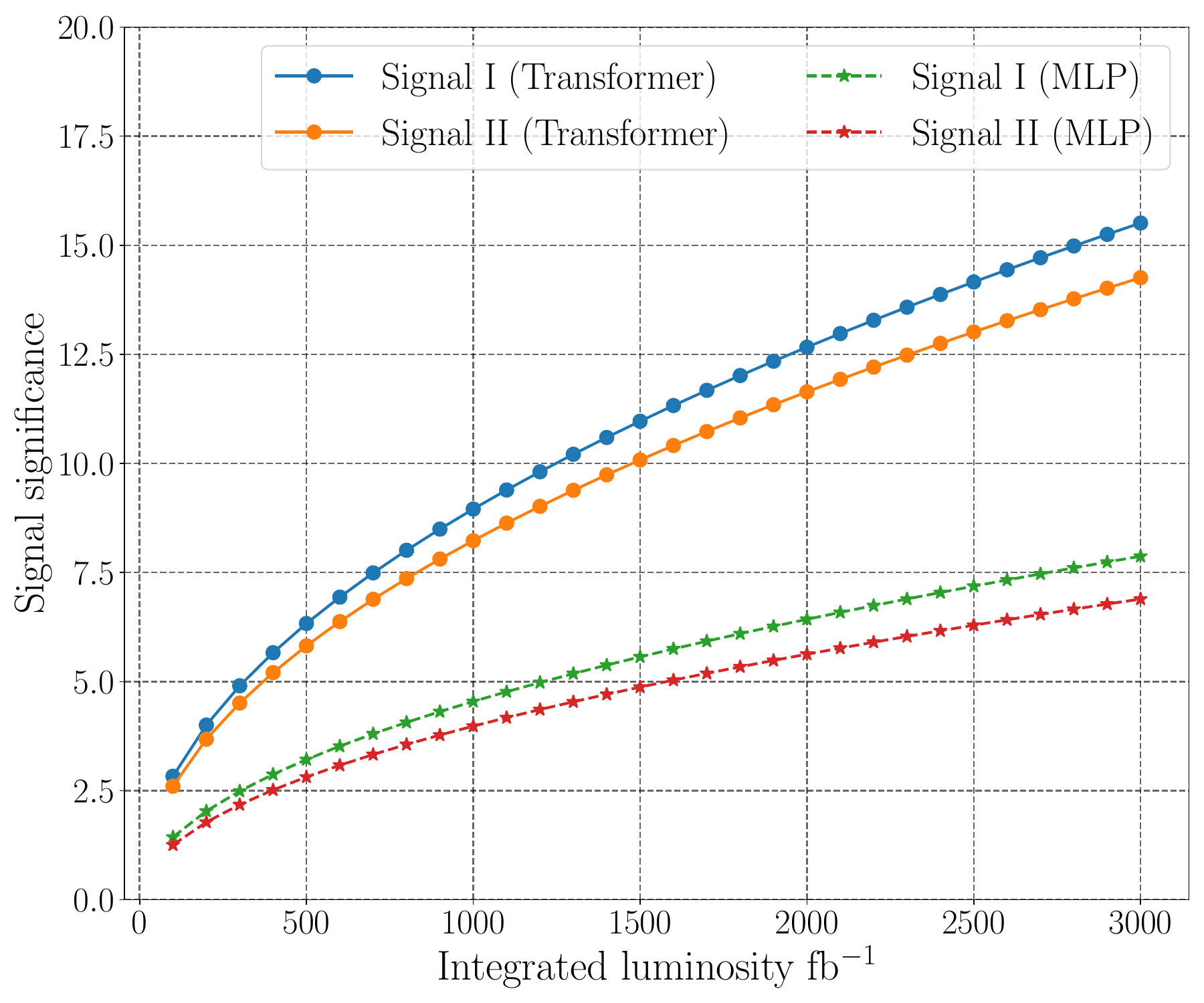}
    \caption{Signal significance as a function of the integrated luminosity. Left: BP1, and right: BP2, with Transformer results shown in solid lines and MLP results in dashed lines.}
    \label{fig:result}
\end{figure*}

Optimization of the signal significance, as a function of the NN output, is performed by varying the cut on the NN score to maximize the signal-to-background yield. For BP1, the optimized Transformer cuts correspond to True Positive Rates (TPRs) of 0.22 and 0.25 for Signal I and Signal II, respectively, while the optimized MLP cuts yield TPR values of 0.16 and 0.21, correspondingly.
For BP2, the optimal Transformer TPR values are 0.31 and 0.28 for Signal I and Signal II, respectively, whereas the MLP achieves TPR values of 0.22 and 0.18, respectively.

Once the cut on the NN output is considered, we compute the signal and background events at a certain integrated luminosity as well as  the signal significance using the following formula \cite{LHCDarkMatterWorkingGroup:2018ufk,Antusch:2018bgr,Antusch:2019eiz}:
\begin{equation}
\sigma = \left[ 2\left( (N_s+N_b)\ln\frac{(N_s+N_b)(N_b+\sigma^2_b)}{N_b^2+(N_s+N_b)\sigma^2_b}\right. \right. -   \left.\left. \frac{N^2_b}{\sigma^2_b}\ln(1+\frac{\sigma^2_b N_s}{N_b(N_b+\sigma^2_b)})         \right) \right]^{1/2}\,,
\end{equation}
with $N_s$, $N_b$ being the number of signal and background events, respectively, and $\sigma_b$ parameterizing the systematic  uncertainty on the latter.  

Figure~\ref{fig:result} shows the signal significance as a function of the integrated luminosity, ranging from $100\,\mathrm{fb}^{-1}$ (approximately Run 2) to $3000\,\mathrm{fb}^{-1}$ (the expected HL-LHC value). The left panel corresponds to BP1 while the right panel corresponds to BP2. For both BPs, the Transformer achieves a substantial improvement over the MLP in significance, a factor of 2, indicating that a potential discovery at the CERN machine may be achievable much sooner in the presence of the former ML environment as opposed to the latter. Furthermore, notice that the  separation  between the two sets of curves in each case (Transformer and MLP) indicates that training on the wrong MC data (Signal II) will induce a clear bias in the real data analysis with respect to the true results (Signal I). Finally, the Transformer appears to better the MLP at controlling this effect  as,
on the one hand, the corresponding spread between the two curves is less (BP1) or comparable (BP2) and, on  the other hand, such a bias is always in the same direction ({\it i.e.}, the color ordering in the solid lines does not change with BP unlike the case of the dashed lines). 
\section{Conclusions}

In summary, we have shown that there is significant scope  in assisting current LHC analyses aimed at isolating SM-like di-Higgs signals ($h_{\rm SM}h_{\rm SM}$) in their dominant decay  mode into four $b$-quarks  when produced resonantly via a heavier CP-even Higgs boson ($H$), crucially including distortion effects of the ensuing BW resonance due to interference effects which are not customarily modelled in experimental analyses and which have been proven to be significant in many theoretical scenarios. This has been based on a sophisticated MC study emulating as close as possible the experimental conditions existing in
the LHC multi-purpose experiments. A combination of preselection cuts, aimed at defining datasets
enriched by signal events without excessively sculpting the relevant backgrounds, and  advanced ML architectures,  either an MLP or a Transformer, enabled us to achieve a twofold result. On the one hand, the advocated preselection helps to improve the signal-to-background ratio in resonant searches beyond the current state-of-the-art while, on the other hand, the ML environment renders such searches less sensitive to the aforementioned interference effects even if the algorithm used was trained on MC datasets not including the latter, with the Transformer narrowly outperforming the MLP in classification tasks.

In order to illustrate the above, we have adopted BPs from the NMSSM,
which corresponding signals in the process $gg\to H\to h_{\rm SM}h_{\rm SM}\to b\bar b b\bar b$
had been previously shown to be greatly affected by a variety of distortion effects due to interferences between the resonant $s$-channel Higgs diagram and the SM ones and to a lesser extent also with diagrams involving light stops as well as possibly another Higgs state propagating in a similar Feynman topology. For such BPs, the ensuing signature is constituted by two fat $b$-jets, inevitably accompanied by some additional hadronic activity. Our analysis proved stable against the jet definition procedure and effective in extracting such signature whichever the kinematic characteristics of the BPs  used.

We therefore advocate the deployment of our approach in actual experimental searches at  the LHC, so as to enable one to thoroughly test a variety of theoretical scenarios in the quest to extracting the shape  of the underlying scalar potential from di-Higgs analyses, as the process tackled here  is the dominant one  at the LHC.

\section*{Acknowledgments}
SM is supported in part through the NExT Institute and STFC Consolidated Grant ST/X000583/1. HW is supported by the Ruth and Nils-Erik Stenb\"ack's Foundation. AH is funded by grant number 22H05113, “Foundation of Machine Learning Physics”, Grant in Aid for Transformative Research Areas and 22K03626, Grant-in-Aid for Scientific Research (C). AH is partially supported by the Science, Technology and Innovation Funding Authority (STDF) under grant number 50806. 

\bibliographystyle{JHEP}
\bibliography{biblo}
\end{document}